\pdfoutput=1

\documentclass{article}

\PassOptionsToPackage{numbers, compress}{natbib}

\usepackage[final,preprint]{neurips_2022}

\usepackage[utf8]{inputenc} 
\usepackage[T1]{fontenc}    
\usepackage{hyperref}       
\usepackage{url}            
\usepackage{booktabs}       
\usepackage{amsfonts}       
\usepackage{nicefrac}       
\usepackage{microtype}      
\usepackage{xcolor}         
\usepackage{graphicx}
\usepackage[font=small,labelfont=bf]{caption}

\title{The Need for Medically Aware Video Compression in Gastroenterology}

\author{%
  Joel Shor\thanks{Authors contributed equally.} \\
  Verily Life Sciences \\
  joelshor@verily.com \\
  \And
  Nick Johnston$^{*}$ \\
  Google Research\\
  nickj@google.com
}

\begin{document}

\maketitle

\begin{abstract}

Compression is essential to storing and transmitting medical videos, but the effect of compression on downstream medical tasks is often ignored. Furthermore, systems in practice rely on standard video codecs, which naively allocate bits between medically relevant frames or parts of frames. In this work, we present an empirical study of some deficiencies of classical codecs on gastroenterology videos, and motivate our ongoing work to train a learned compression model for colonoscopy videos.
We show that \textbf{two of the most common classical codecs, H264 and HEVC, compress medically relevant frames statistically significantly worse than medically nonrelevant ones}, and that polyp detector performance degrades rapidly as compression increases. We explain how a learned compressor could allocate bits to important regions and allow detection performance to degrade more gracefully. Many of our proposed techniques generalize to medical video domains beyond gastroenterology.
\end{abstract}

\section{Introduction}

Colorectal cancer is the third most common cancer diagnosed in the US~\cite{colorectal_cancer_us} and worldwide~\cite{colorectal_cancer_world}. Colonoscopies, a video-based diagnostic procedure, are one of the most common screening tools. 
Over 15 million colonoscopies are performed in the US each year~\cite{us_colonoscopies}, leading to an enormous amount of video transmission and storage 
for tasks like medical records, physician education, report generation, and training medical models. Data-driven machine learning models that perform polyp detection~\cite{polyp_detection} and coverage detection~\cite{coverage} need these videos to train and evaluate.

H264~\cite{h264} and High Efficiency Video Coding (HEVC)~\cite{hevc} are two of the most common classical (non-machine learning based) video codecs. The quality versus size tradeoff is controlled using the "Quantization Parameter" (QP), which ranges from 0 (lossless) to 51 (most compressed). QP and other parameters can be tuned to match target quality requirements, but the transform-based compression algorithms are unable to take advantage of video domain specific properties, such as the texture and camera motion that are specific to colonoscopies. 

Previous attempts to achieve diagnostically-lossless compression, or to reduce the degradation of medically relevant regions of interest, include modifying classical codecs to take advantage of medical properties~\cite{medical_video_compression1, medical_video_compression3}, leveraging superresolution~\cite{medical_video_compression2}, and applying different quality compression algorithms on specifically identified regions of interest~\cite{medical_image1}. Researchers have also explored tuning classical codecs to specific medical domains, such as ultrasound~\cite{ultrasound_compression}. However, fully data-driven, modern image and video compression algorithms~\cite{learned_video_compression1, toderici2017full} have yet to be applied to the medical domain.



\begin{figure}
  \centering
  \includegraphics[height=0.21\textwidth]{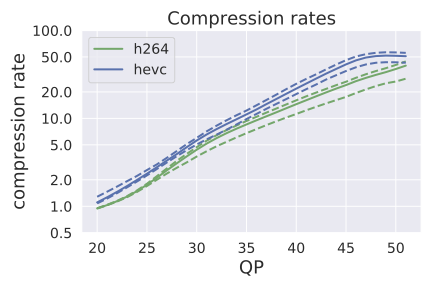}
  \includegraphics[height=0.21\textwidth]{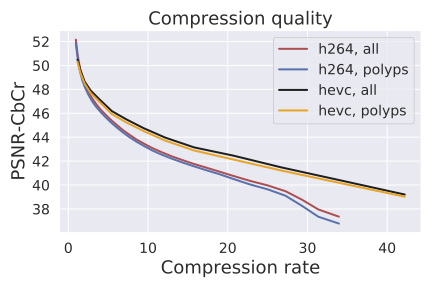}
  \includegraphics[height=0.21\textwidth]{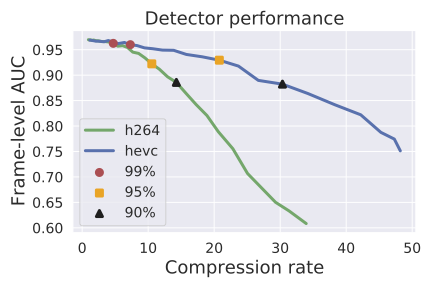} 
  \caption{\textbf{Left)} 25th, 50th, and 75th percentile compression rates across 80 colonoscopy videos for difference QP values of H264 and HEVC. \textbf{Middle)} Frame compression frame quality, as measured by PSNR-CbCr. See Appendix for other metrics. \textbf{Right)} Polyp detector performance as a function of compression factor.}
  \label{fig:results}
\end{figure}

\section{Experiments}

\textbf{Data:} As in \cite{polyp_detection}, our colonoscopy video dataset was collected from screening colonoscopy procedures performed in three hospitals. We used a detector trained on a 16K video subset of the dataset, and we computed our classical codec analyses on a different 80 video subset (2.2M frames, 15K polyp frames). All videos and metadata were deidentified according to the Health Insurance Portability and Accountability Act Safe Harbor. Ground truth polyp labeling was provided by the gastroenterologist annotators described in \cite{polyp_detection}.
The annotators were paid on an hourly basis, and pay was not based on the results they provided. The videos were compressed using H264 QP20 when transmitted from the hospitals. To justify our analyzing already-compressed videos, we used a small number of lossless videos to investigate the impact of re-compressing compressed videos to QP$N$ (so called ``generation loss"). We found that the impact was negligible (see Appendix for details).

\textbf{Compression metrics:} We evaluate frame quality using two standard image metrics: PSNR-CbCr and PSNR-Y. PSNR is the standard "Peak Signal to Noise Ratio" derived from the mean squared error between pixels in the original frame and the compressed frame. The value of the pixels depends on the type of PSNR computation:``CbCr" corresponds to PSNR between chroma of the frames, and ``Y" corresponds to PSNR between luminance of the frames.

\textbf{Polyp detector metrics:} We evaluate polyp detector performance using the same methodology as \cite{polyp_detection}. We report the AUC value of sensitivity versus false positive rate for various thresholds of the detector. The threshold determines how "confident" the polyp detector must be to register a detection, so the AUC curve captures the detector's performance across a range of sensitivity scores.

\textbf{Polyp detector model:} The polyp detector is a production-grade polyp detector, as described in \cite{polyp_detection}. The architecture is RetinaNet~\cite{retinanet} with training and hyperparameters described in \cite{polyp_detection}. The polyp detector demonstrates state-of-the-art performance on colonoscopy procedure videos as well as diagnostically challenging polyps.

\begin{figure}
  \centering
\includegraphics[width=0.4\textwidth]{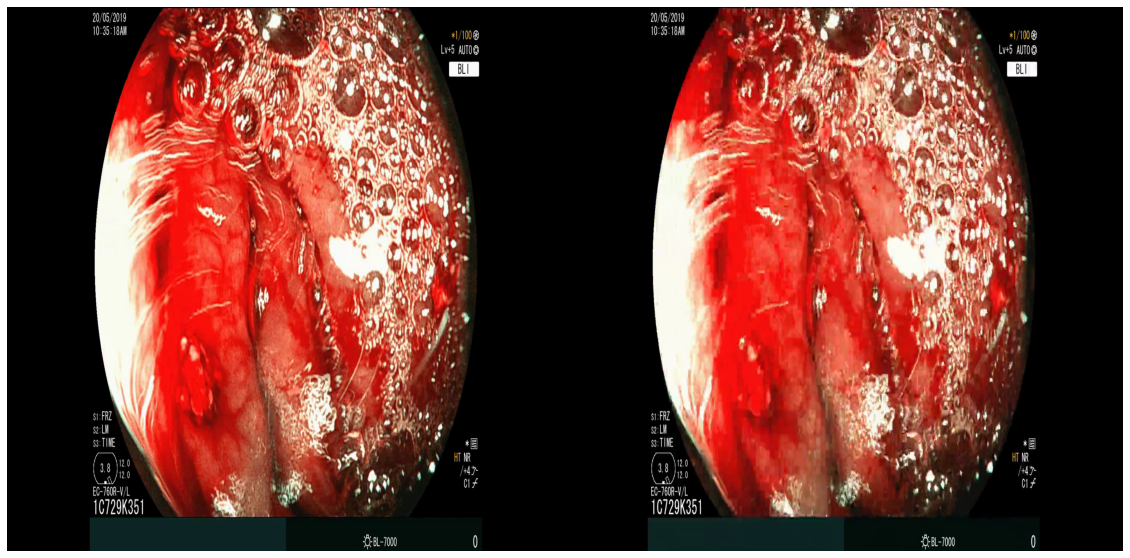}
\includegraphics[width=0.4\textwidth]{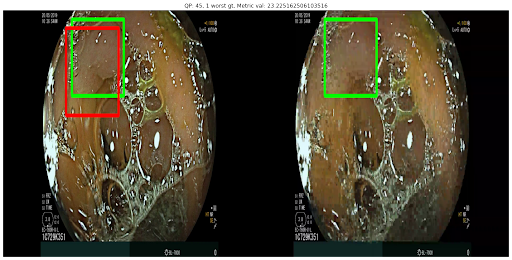} \\
\includegraphics[width=0.4\textwidth]{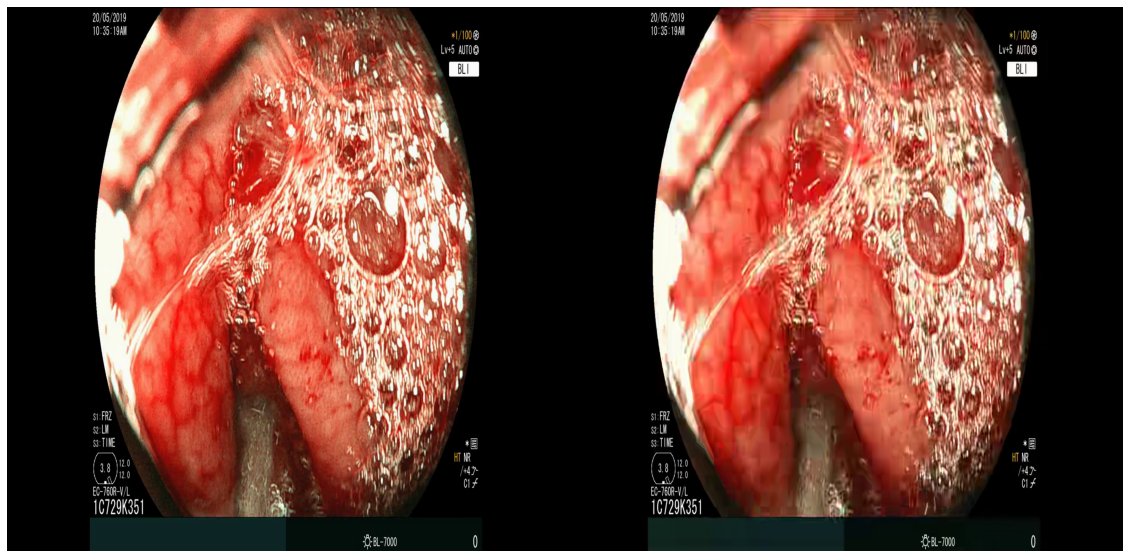}
\includegraphics[width=0.4\textwidth]{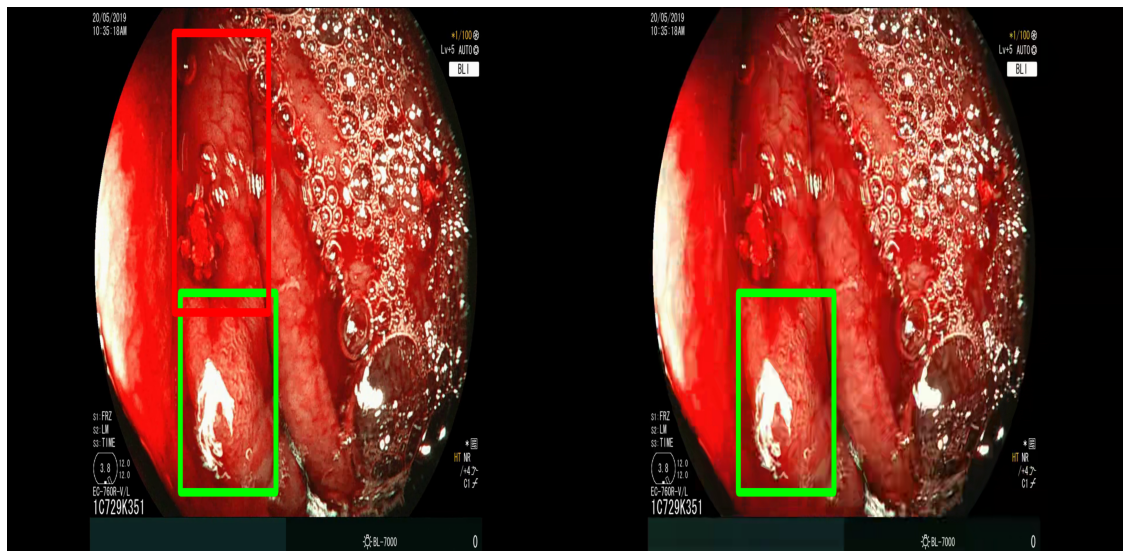} \\
\includegraphics[width=0.6\textwidth]{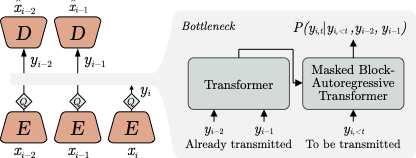}
\caption{Qualities measured by PSNR-CbCr. Concatenated images are original frame on the left, compressed on the right. \textbf{Left upper (lower))} Worst compressed frames by H264 (HEVC) QP40. \textbf{Right upper (lower))} Worst compressed frames by H264 (HEVC) QP40 that have a polyp. \textbf{3rd row)} A high level description of learned neural video compression (diagram from \cite{vct}).}
\label{fig:worst_compressed}
\end{figure}

\section{Results}
\textbf{Compression rates}: Figure \ref{fig:results} (left) shows the distribution of compression rates on colonoscopy videos for H264 and HEVC by QP value. At QP 51 (the highest compression rate for both classical codecs), HEVC achieved significantly more compression: the 25th, 50th, and 75th percentile compression rates were (43.6, 51.8, 56.5) for HEVC and (28.3, 39.8, 44.5) for H264. 

\textbf{Compression quality}: Figure \ref{fig:results} (middle) shows the compression rate versus frame quality distribution for H264 and HEVC.  Importantly, we see that \textbf{H264 and HEVC compress the most medically relvant frames statistically significantly worse}: treating each QP value separately, a two-sided Kolmogorov-Smirnov test between distribution of PSNR-CbCr shows that the frame quality is lower for polyp frames than for all frames. For each QP value, $N_1=2189948$, $N_2=15457$, H264 (HEVC) maximum p-value over all tests is $1.4*10^{-118}$ ($1.4*10^{-30}$), mean test statistic $0.13$ ($0.11$). For the same test with PSNR-Y, see the Appendix. Figure \ref{fig:worst_compressed} top two rows show the lowest quality compressed frames inside the body according to PSNR-CbCr, with and without polyps (for the absolute worst quality compressed frames, see the Appendix).

\textbf{Detector performance}: Figure \ref{fig:results} (right) shows the polyp detector performance as a function of compression rate. Videos can be compressed by factors of 4.7x and 7.3x before dropping below 99\% the base performance for H264 and HEVC respectively, 10.6x and 20.8x for 95\%, and 14.3x and 30.4x for 90\%. In addition to getting better frame quality and higher compression rates, \textbf{the detector performs 0.057 AUC better and a 29\% relative improvement\footnote{relative improvement defined as $(AUC_{hevc} - AUC_{h264}) / (1.0 - AUC_{h264})$} on HEVC videos compared to H264, as the same compression rate}. The same holds in the ``practical" regime of compression rates that achieve at least 95\% the AUC of the original model: AUC is on average 0.26 AUC improved with a 23\% relative improvement in AUC. 

\section{Future work}

Ongoing work involves addressing the deficiencies of classical codecs in preserving medically relevant information. Our primary proposed solution is to leverage data-driven or ``learned" compression using the recent Video Compression Transformer model~\cite{vct}. This model shows especially useful properties, such as the ability to capture domain-specific synthetic camera motion and domain-specific texture characteristics. We also plan to explore the complex relationship between training on videos compressed using one algorithm (e.g. H264 QP20), but run on a different type of compression during inference (e.g. lossless).

There are a number of ways to inject colonoscopy-specific information into a learned compressor. First, we can simply train the compression model on colonoscopy data, and it will learn to recreate videos with colonoscopy texture and video motion. Second, we can use colonoscopy data as a validation set to pick model hyperparameters (see \cite{shor2022universal} for an example of the potential impact of just using data to select model hyperparameters). Third, we can explore oversampling from polyp frames during training. Fourth, we can add a weight factor to the training loss on polyp frames during training. Fifth, we can add a weight factor on the training loss for polyp subregions of frames. Sixth, we can add polyp detectors directly to the compressor loss function (as well as other differentiable medical models, such as polyp type classification). Seventh, we can co-train the compression model with the detection model to maximize detector performance on the compressed videos.



\begin{ack}

\section{Acknowledgements}

This work was funded by Verily LLC. We'd like to thank Roman Goldberg and Ehud Rivlen for their technical guidance. We'd like to thank Joe Shao, Stephen Lanham, Bimba Rao, Josh Widen, Bryce Evans, and Brijesh Patel for their suggestions and feedback.
\end{ack}

{
\small
\bibliographystyle{plain}
\bibliography{references}
}

\iftrue
\section*{Checklist}

The checklist follows the references.  Please
read the checklist guidelines carefully for information on how to answer these
questions.  For each question, change the default \answerTODO{} to \answerYes{},
\answerNo{}, or \answerNA{}.  You are strongly encouraged to include a {\bf
justification to your answer}, either by referencing the appropriate section of
your paper or providing a brief inline description.  For example:
\begin{itemize}
  \item Did you include the license to the code and datasets? \answerNA{Code not released.}
  \item Did you include the license to the code and datasets? \answerNo{The code and the data are proprietary.}
  \item Did you include the license to the code and datasets? \answerNA{}
\end{itemize}
Please do not modify the questions and only use the provided macros for your
answers.  Note that the Checklist section does not count towards the page
limit.  In your paper, please delete this instructions block and only keep the
Checklist section heading above along with the questions/answers below.

\begin{enumerate}

\item For all authors...
\begin{enumerate}
  \item Do the main claims made in the abstract and introduction accurately reflect the paper's contributions and scope?
    \answerYes{See Experiments and Results section for justification.}
  \item Did you describe the limitations of your work?
    \answerYes{}
  \item Did you discuss any potential negative societal impacts of your work?
    \answerYes{}
  \item Have you read the ethics review guidelines and ensured that your paper conforms to them?
    \answerYes{}
\end{enumerate}

\item If you are including theoretical results...
\begin{enumerate}
  \item Did you state the full set of assumptions of all theoretical results?
    \answerNA{No theoretical results.}
        \item Did you include complete proofs of all theoretical results?
    \answerNA{No proofs.}
\end{enumerate}

\item If you ran experiments...
\begin{enumerate}
  \item Did you include the code, data, and instructions needed to reproduce the main experimental results (either in the supplemental material or as a URL)?
    \answerNo{}
  \item Did you specify all the training details (e.g., data splits, hyperparameters, how they were chosen)?
    \answerNA{No new models trained.}
        \item Did you report error bars (e.g., with respect to the random seed after running experiments multiple times)?
    \answerYes{We show error bars in all of our analyses.}
        \item Did you include the total amount of compute and the type of resources used (e.g., type of GPUs, internal cluster, or cloud provider)?
    \answerNA{No new models trained.}
\end{enumerate}

\item If you are using existing assets (e.g., code, data, models) or curating/releasing new assets...
\begin{enumerate}
  \item If your work uses existing assets, did you cite the creators?
    \answerYes{We cite the polyp detection paper for their models.}
  \item Did you mention the license of the assets?
    \answerYes{Data is not publically available.}
  \item Did you include any new assets either in the supplemental material or as a URL?
    \answerNA{No new assets.}
  \item Did you discuss whether and how consent was obtained from people whose data you're using/curating?
    \answerYes{All patient data was provided after written consent.}
  \item Did you discuss whether the data you are using/curating contains personally identifiable information or offensive content?
    \answerYes{Data has been anonymized according to HIPAA.}
\end{enumerate}

\item If you used crowdsourcing or conducted research with human subjects...
\begin{enumerate}
  \item Did you include the full text of instructions given to participants and screenshots, if applicable?
    \answerNA{}
  \item Did you describe any potential participant risks, with links to Institutional Review Board (IRB) approvals, if applicable?
    \answerNA{}{}
  \item Did you include the estimated hourly wage paid to participants and the total amount spent on participant compensation?
    \answerNo{Participants were paid an hourly wage with no contingency on performance. The wage information for these participants is not publicly available.}
\end{enumerate}

\end{enumerate}


\section{Potential negative societal impact}

Since this research deals with medical data and has clinical applications, the usual potential negative societal impacts are applicable. Incorrect model training or deidentification can lead to accidentally leaked and/or memorized patient information. A model with biased performance can lead to negative clinical outcomes for protected subgroups. Finally, good performance on a test set doesn't mean the model is necessarily ready for immediate clinical application, and having a performant model doesn't necessarily mean it can replace a medical professional.


\appendix

\section{Appendix}
\subsection{Generation loss}

Since our dataset was compressed to be transferred from hospitals, we first investigated the impact of multiple stages of compression ("generation loss"). We used lossless video (24 seconds, 1.4GB) collected from an endoscope viewing dyed, non-human tissue. This gave the video the motion and texture characteristics of a colonoscopy. We then compared video quality between two compression schemes: compressing using H264 QP $N$, where $20 <= N <= 51$, and compressing to H264 QP20, then to H264 QP$N$ (when referring explicitly to this comparison, we will concisely refer to the latter as 'QP'). We then compared the average frame quality between these two schemes using two quality metrics (see 'Metrics' section). The quality differences between these two compression schemes was minor in terms of PSNR-CbCr, which justifies our working with video data already compressed by H264 QP20.

\begin{figure}
  \centering
  \includegraphics[width=0.32\textwidth]{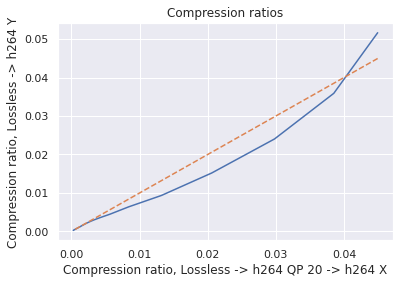}
  \includegraphics[width=0.32\textwidth]{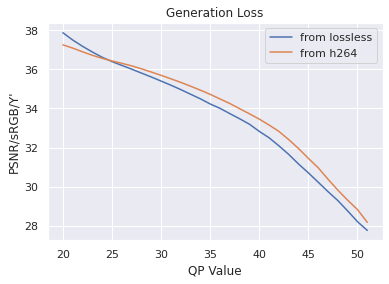}
  \includegraphics[width=0.32\textwidth]{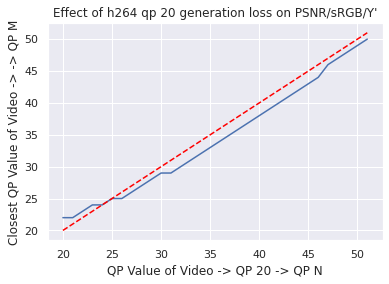}
   \\
  \includegraphics[width=0.32\textwidth]{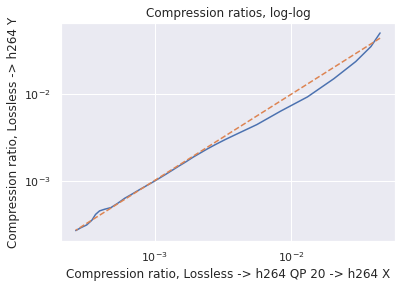}
  \includegraphics[width=0.32\textwidth]{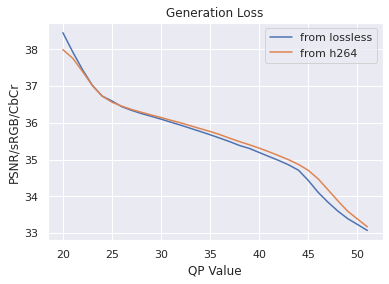}
  \includegraphics[width=0.32\textwidth]{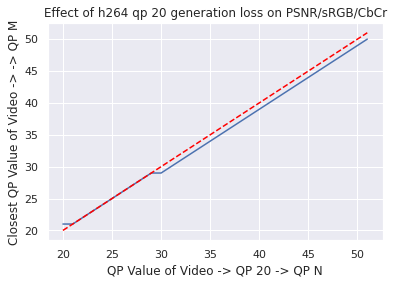}
  
  \caption{\textbf{Left upper)} Comparison of compression rates for two compression methods using the same QP value $N$. y-axis is the compression rate of compressing from lossless directly to H264 QP$N$. x-axis is the compression rate of compressing to H264 QP$N$ through H264 QP20. Dotted line is the $x=y$. Note that for lower compression rates (lower QP value), compressing through H264 QP20 gives a smaller file sizes. \textbf{Left lower)} Same as (left upper), but a log-log plot. \textbf{Middle upper)} Frame quality metric PSNR-Y, averaged across compressed frames, for the two methods of compression. \textbf{Middle upper)} Same as (middle upper) but for PSNR-CbCr. \textbf{Right upper)} Correspondence between QP values of the two compression methods, according to closest PSNR-Y values. \textbf{Right lower)} Same as Right upper, but for PSNR-CbCr.}
\end{figure}

\subsection{Compression quality}

See Figure \ref{fig:qual} for compression quality as a function of QP value instead of compression rate, as well as quality measured by PSNR-Y. Furthermore, the Kolmogorov-Smirnov tests on PSNR-Y show the same behavior: treating each QP value separately, a two-sided  between distributions shows that the frame quality is lower for polyp frames than for all frames (for each QP value, $N_1=2189948$, $N_2=15457$, H264 PSNR-Y max p-value is $9.5*10^{-77}$, mean test statistic $0.10$, HEVC PSNR-Y maximum p-value over all tests is $2.4*10^{-13}$, mean test statistic $0.13$). This result holds for each QP value between 20 and 51.

\begin{figure}
  \centering
  \includegraphics[width=0.32\textwidth]{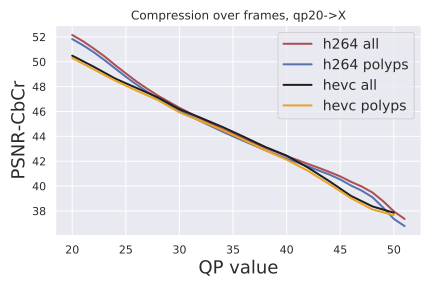}
  \includegraphics[width=0.32\textwidth]{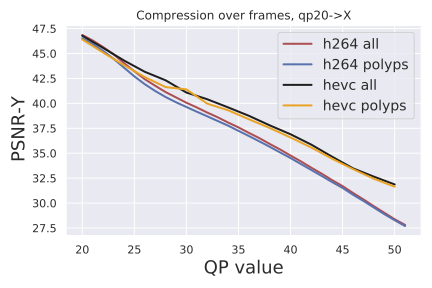}
  \includegraphics[width=0.32\textwidth]{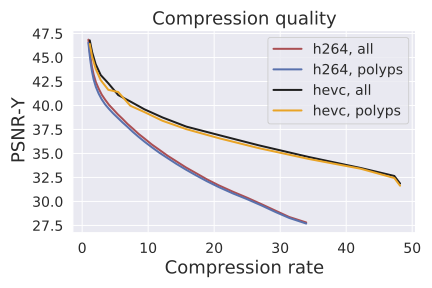}
  
  \caption{Quality vs QP value and quality vs compression for different quality metrics.}
  \label{fig:qual}
\end{figure}

\subsection{Lowest quality compressed frames}

Figure \ref{fig:worst_compressed} shows the worst compressed colonoscopy frames inside the body, and with polyps. Interestingly, the worst quality compressed frames were actually outside the body. We include them here for completeness.

\begin{figure}
  \centering
  \includegraphics[width=0.49\textwidth]{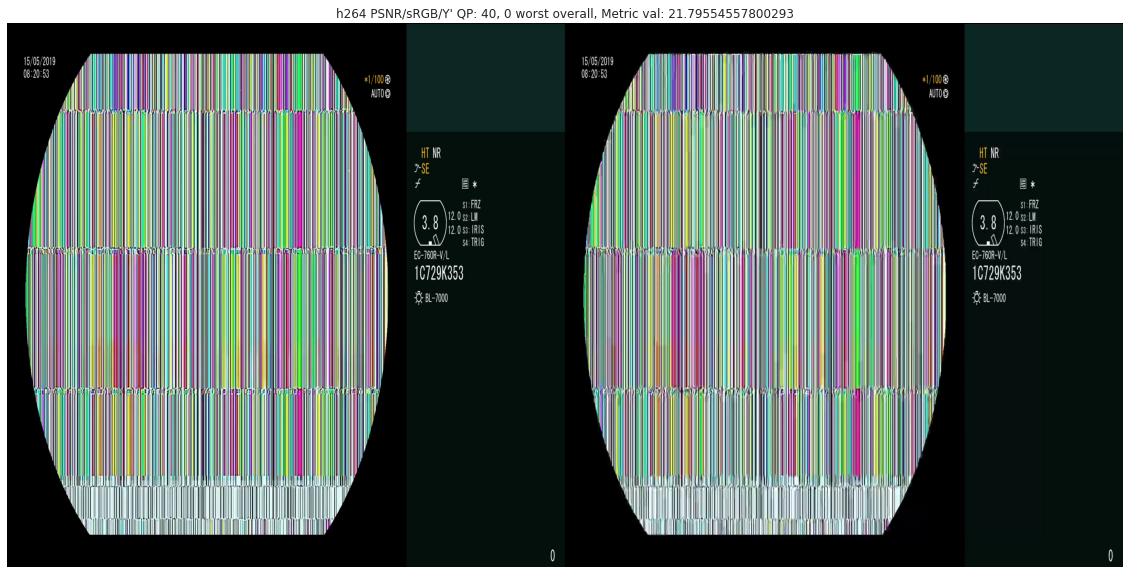}
  \includegraphics[width=0.49\textwidth]{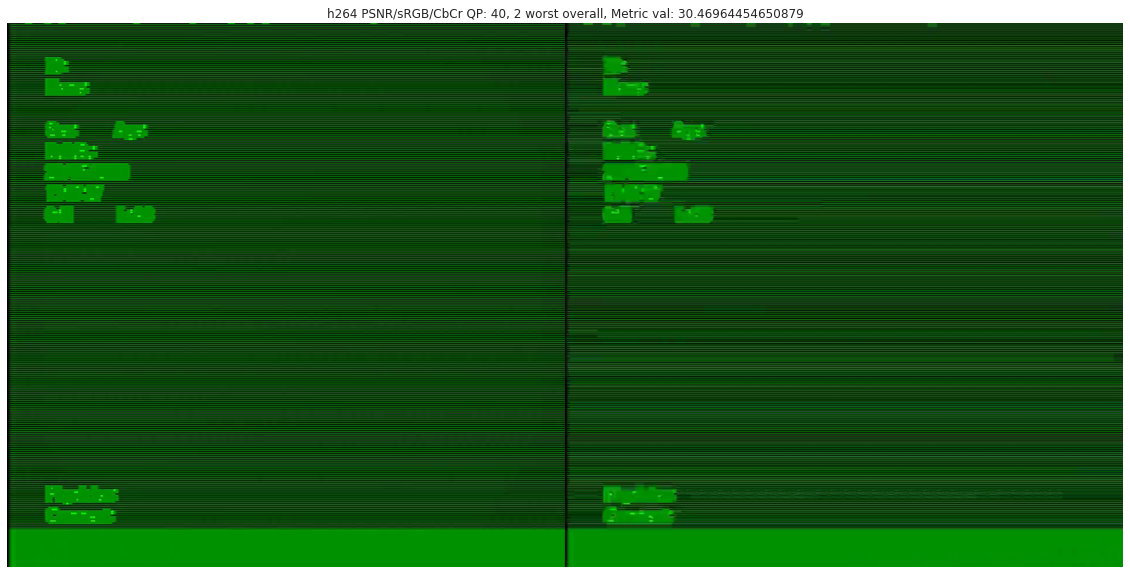}
  
  \caption{Overall frames with the lowest quality compression for H264 QP40.}
\end{figure}
\fi

\end{document}